\documentclass[5p,twocolumn,sort&compress,times,preprint,normalem]{elsarticle}

\usepackage{graphics}
\usepackage{graphicx}
\usepackage{amsfonts, amsthm, amssymb}
\usepackage{color}
\usepackage{ascmac}
\usepackage{url}
\usepackage{ulem}
\usepackage{bm}
\usepackage{bbold}
\usepackage{amsmath}
\usepackage{algorithm}
\usepackage{algpseudocode}
\usepackage{braket}
\usepackage{multirow}
\usepackage{subfiles}
\usepackage{cprotect}
\usepackage{epsfig}
\usepackage{xspace}
\usepackage{tipa}
\usepackage[final]{listings}

\newcommand{\red}[1]{#1}
\newcounter{bla}

\journal{Computer Physics Communications}

\begin{document}

\begin{frontmatter}



\title{Data-analysis software framework 2DMAT and its application to experimental measurements for two-dimensional material structures}


\author[a]{Yuichi Motoyama}\ead{y-motoyama@issp.u-tokyo.ac.jp}
\author[a]{Kazuyoshi Yoshimi}\ead{k-yoshimi@issp.u-tokyo.ac.jp}
\author[b]{Izumi Mochizuki}
\author[c]{Harumichi Iwamoto}
\author[d]{Hayato Ichinose}
\author[c,b,d,e]{Takeo Hoshi}\ead{ hoshi@tottori-u.ac.jp}

\address[a]{Institute for Solid State Physics, University of Tokyo, Kashiwa-shi, Chiba 277-8581 Japan}
\address[b]{Slow Positron Facility, Institute of Materials Structure Science, High Energy Accelerator Research Organization (KEK), Oho 1-1, Tsukuba, Ibaraki 305-0801, Japan}
\address[c]{Department of Mechanical and Physical Engineering, Faculty of Engineering, Tottori University,  Tottori-shi, Tottori 680-8552, Japan}
\address[d]{Department of Engineering, Graduate School of Sustainability Science, Tottori University,  Tottori-shi, Tottori 680-8552, Japan}
\address[e]{Advanced Mechanical and Electronic System Research Center, Faculty of Engineering, Tottori University,  Tottori-shi, Tottori 680-8552, Japan}

\begin{abstract}
An open-source data-analysis framework \verb|2DMAT| 
has been developed for experimental measurements
of two-dimensional material structures. 
\verb|2DMAT| offers five analysis methods: 
(i)  Nelder-Mead  optimization,
(ii) grid search, 
(iii) Bayesian optimization, 
(iv) replica exchange Monte Carlo method, and
(v) population-annealing Monte Carlo method. 
Methods (ii) through (v) are implemented by parallel computation,
which is efficient not only for personal computers but also for supercomputers.
The current version of 2DMAT is applicable to total-reflection high-energy positron diffraction (TRHEPD), surface X-ray diffraction (SXRD), and low-energy electron diffraction (LEED) experiments by installing corresponding forward problem solvers that generate diffraction intensity data from a given dataset of the atomic positions.
The analysis methods are general and can be applied also to other experiments and problems. 
\end{abstract}

\begin{keyword}
data analysis for experimental measurements,
Nelder-Mead optimization, 
Bayesian optimization, 
grid search, 
replica exchange Monte Carlo method, 
population-annealing Monte Carlo method,
two-dimensional material, 
total-reflection high-energy positron diffraction,
surface X-ray diffraction,
low-energy electron diffraction
\end{keyword}

\end{frontmatter}

{\bf PROGRAM SUMMARY}
  %Delete as appropriate.

\begin{small}
\noindent
2DMAT-- Open-source data-analysis framework for experimental measurement 
of two-dimensional material structures \\
{\em Authors:} Yuichi Motoyama, Kazuyoshi Yoshimi, 
Izumi Mochizuki, Harumichi Iwamoto, Hayato Ichinose,  Takeo Hoshi
\\
{\em Program title:} 2DMAT\\
{\em Journal reference:}   \\
  %Leave blank, supplied by Elsevier.
{\em Catalogue identifier:}                                   \\
  %Leave blank, supplied by Elsevier.
{\em Program summary URL:} \\
https://www.pasums.issp.u-tokyo.ac.jp/2dmat/ \\
{\em Licensing provisions:} GNU General Public License v3.0\\
  %enter "none" if CPC non-profit use license is sufficient.
{\em Programming language:} Python~3                                   \\
{\em Computer:} Any architecture\\
  %Computer(s) for which program has been designed.
{\em Operating system:} Unix, Linux, macOS  \\
  %Operating system(s) for which program has been designed.
{\em RAM:} Depends on the number of variables \\
{\em Number of processors used:} Arbitrary \\
%If more than one processor.
{\em Keywords: Nelder-Mead optimization, 
Bayesian optimization, 
grid-based search, 
replica exchange Monte Carlo method, 
population-annealing Monte Carlo method,
total-reflection high-energy positron diffraction (TRHEPD),
surface X-ray diffraction (SXRD),
low-energy electron diffraction (LEED).
}
\\
  % Please give some freely chosen keywords that we can use in a
  % cumulative keyword index.
%{\em Classification:X.X XXXXXX }
  %Classify using CPC Program Library Subject Index, see (
  % http://cpc.cs.qub.ac.uk/subjectIndex/SUBJECT_index.html)
%e.g. 4.4 Feynman diagrams, 5 Computer Algebra.
%\\
{\em External routines/libraries:} Numpy, Scipy, Tomli, mpy4py \\
{\em Nature of problem:}
Analysis of experimental measurement data\\
{\em Solution method:} Optimization, grid-based global search, Monte Carlo method\\
%{\em Unusual features:}\\
%\\
{\em Code Ocean capsule:} (to be added by Technical Editor) 
\\
%{\em Running time:}\\
%Give an indication of the typical running time here.
\\

%\begin{thebibliography}{0}
%\bibitem{1}Reference 1         % This list should only contain those items referenced in the                 
%\bibitem{2}Reference 2         % Program Summary section.   
%\bibitem{3}Reference 3         % Type references in text as [1], [2], etc.
%                               % This list is different from the bibliography at the end of 
%                               % the Long Write-Up.
%\end{thebibliography}
%* Items marked with an asterisk are only required for new versions
%of programs previously published in the CPC Program Library.\\
\end{small}

%%%%%%%%%%%%%%%%%%%%%%%%%%%%%%%%%%%%%%
\section{Introduction \label{SEC-INTRO}}
One major issue in computational physics is to construct a
fast and reliable data-analysis method for experimental measurements.
The data-analysis procedure is to obtain 
the target physical quantity
$X \equiv (X_1, X_2, \dots, X_n)$ 
from the experimentally observed data $D \equiv (D_1,D_2,\dots,D_m)$. 
A typical case is the inverse problem, in which  
%the experimentally observed data 
$D$ is caused by %the quantity 
$X$ and is written as a function of $X$ ($D=D_\text{cal}(X)$), which is referred to as a forward problem solver. 
The inverse problem is solved, in principle, by searching for the minimum of the objective function
\begin{equation}
F = F(X) \equiv d(D_{\rm cal}(X),  D),
\label{EQ-RESIDUAL-FUNC}
\end{equation}
where $d(u, v)$ denotes a distance function between the two vectors $u=(u_1,\dots,u_m)$ and $v=(v_1,\dots,v_m)$. Hereinafter the 2-norm form 
\begin{equation}
d(u,  v) \equiv \left(\sum_{i=1}^m (u_i-v_i)^2 \right)^{1/2}, 
\label{EQ-RESIDUAL-FUNC-2norm}
\end{equation}
is used, 
except where indicated. 
The objective function $F(X)$ may have local minima and may be affected by 
uncertainties due to the measurement conditions and the apparatus.

Here, we focus on quantum-beam diffraction experiments
for the structure determination of two-dimensional (2D) materials, 
since 2D materials hold great promise for industrial applications, such as next-generation electronic devices, and catalysts. 
\red{Nowadays, the structure of 2D materials are measured by total-reflection high-energy positron diffraction (TRHEPD) \cite{HUGENSCHMIDT_2016_SurfSciRep_rev, Fukaya_2018_JPHYSD, Fukaya_2019_Book_SurfaceStructure}, surface X-ray diffraction (SXRD)  ~\cite{FEIDENHANSL_1989_SXRD_review,Tajiri_2020_SXRD_review} and low-energy electron diffraction (LEED) \cite{VANHOVE-1993}. 
In these experiments, $D$ is the diffraction intensity and $X$ is the atom position in two-dimensional materials. However, 2D materials are generally more difficult to measure the structure (atom position) than three-dimensional materials, since the diffraction beam intensity of 2D materials is weak, and the types of structures are diverse.}

\red{
The present paper reports that we have recently developed} \verb|2DMAT|, \red{
an open-source data-analysis framework for two-dimensional material structures.
The data-analysis methods are based on the above inverse-problem approach 
for TRHEPD, SXRD, and LEED experiments.
In addition,} \verb|2DMAT| \red{is applicable to other problems if the user prepares an objective function $F=F(X)$.  
The software framework features parallel algorithms using the Message Passing Interface (MPI) for efficient data analysis not only by workstations with many-core CPU's but also by supercomputers. An early-stage package was developed as a private program 
\cite{TANAKA2020_ACTA_PHYS_POLO, TANAKA_2020_Preprint,HOSHI_2021_SION}.
We then added several new features to the code and released the present package as an open-source software framework.}

The remainder of the present paper is organized as follows. 
Section \ref{SEC-ALGORITHM} gives 
an overview of the algorithms for the inverse problem.
Section \ref{SEC-SOLVER} explains 
the supported experiments and problems: TRHEPD, SXRD, LEED, and others. 
Section \ref{SEC-SOFT} explains the present software framework. 
Section \ref{SEC-EXAMPLES} is devoted to examples.
Finally, we summarize the present paper \red{and comment on future utilities} in Section \ref{SEC-SUMMARY}.

%%%%%%%%%%%%%%%%%%%%%%%%%%%%%%%%%%%%%%
\section{Analysis algorithms for the inverse problem \label{SEC-ALGORITHM}}

This section is devoted to an overview of the following five analysis algorithms implemented in \verb|2DMAT|:
(i)  Nelder-Mead method~\cite{NMmethod-a, NMmethod-b},
(ii) grid search, 
(iii) Bayesian optimization (BO) method~\cite{GPML2005}, 
(iv) replica exchange MC (REMC) method~\cite{REMC-PAPER}, and
(v) population-annealing MC (PAMC) method~\cite{PAMC-PAPER}. 
These algorithms commonly use computation of the objective function $F=F(X)$ in $n$-dimensional parameter space ($X \equiv (X_1, \dots, X_n)$).
%The latter four methods use MPI (MPI for Python~\cite{mpi4py2005, mpi4py2008, mpi4py2011, mpi4py2021}) for parallel computation.
For the latter four methods, MPI (MPI for Python~\cite{mpi4py2005, mpi4py2008, mpi4py2011, mpi4py2021}) can be used for parallel computation.

\subsection{Nelder-Mead method \label{SEC-NM-METHOD} }

The Nelder-Mead optimization method, also known as the downhill simplex method\cite{NMmethod-a, NMmethod-b}, is a gradient-free optimization method in which the gradient, $\nabla_X F = (\partial F/ \partial X_1, \partial F/ \partial X_2, \dots, \partial F/ \partial X_n)$, is not required.
The optimal solution is searched for by systematically moving pairs of $n+1$ coordinate points $\{ X^{(l)} \}_{l=1, \dots, n+1}$, according to the value of the objective function $F(X^{(l)})$ at each point $X^{(l)}$.
In \verb|2DMAT|, the function \verb|scipy.optimize.minimize| in scipy~\cite{URL-scipy} is used, and the method is not parallelized. 

%%%%%%%%%%%%%%%%%%%%%%%%%%%%%
\subsection{Grid-search method}

The grid-search method is an algorithm for searching for the minimum value of $F(X)$ by computing $F(X)$ for all of the candidate points 
%$\{ X^{(\text{g} i)} \}_i$ 
in the parameter space prepared in advance. 
In \verb|2DMAT|, the set of candidate points is equally divided and automatically assigned to each process for trivial parallel computation.

\subsection{Bayesian optimization method}

The BO method is a black-box optimization method\cite{GPML2005}.
In BO, a surrogate model with a parameter set $\theta$, $G(X; \theta)$, is trained by an observed dataset for representing $F(X)$.
Then, by using the trained model $G(X; \theta)$, the next candidate $X^*$ to be observed is obtained by optimizing an acquisition function calculated from the expectation value and the variance of $G(X; \theta)$.
For example, the probability of improvement $\text{PI}(X)$, which is the probability that $X$ becomes the new optimal solution when $X$ is observed, is one of the widely-used acquisition functions.
\verb|2DMAT| uses a BO library, \verb|PHYSBO|~\cite{PHYSBO-URL, PHYSBO}.
As for the grid-search method, \verb|PHYSBO| requires the set of candidate points in advance but enables speed-up for yielding the next candidate by trivial parallelization.

\subsection{Replica-exchange Monte Carlo method}

Thus far, we have considered the optimization problem, i.e., the minimization of $F(X; D)$ for determining the optimal value of $X$ from the experimentally observed data $D$.
Hereinafter, we will determine the conditional distribution of $X$ under $D$ (also known as the posterior probability distribution), $\pi(X|D)$.
Then, $\pi(X|D)$ is given by
\begin{equation}
\pi(X|D) = \frac{\pi(D|X) \pi(X)}{Z},
\end{equation}
where $\pi(D|X)$ is the likelihood function, $\pi(X)$ is the prior probability distribution of $X$, and $Z = \int dX \pi(D|X) \pi(X) $ is the normalization factor.
The likelihood $\pi(D|X)$ is defined  as $\pi(D|X) = \exp(-F/\tau)$.
The given parameter $\tau(>0)$ is a measure of the tolerable uncertainty, which is called the \lq temperature' based on the analogy of statistical physics.
A constraint such as $X_i^\text{(lower)} \le X_i \le  X_i^\text{(upper)}$
is reduced to the uniform prior distribution in the region $\pi(X)$.
The normalization factor $Z$, also called the partition function, is useful for choosing the proper model, $D_\text{cal}(X)$.
By using these terms, the posterior probability distribution $\pi(X|D; \tau)$ can be rewritten as
\begin{equation}
\pi(X|D; \tau) = \frac{\exp(-F(X; D)/\tau)}{Z\Omega},
\end{equation}
where $\Omega$ is the volume of the region.
Although it looks simple, it is generally difficult to evaluate $\pi(X|D; \tau)$ for all of the parameter space when the number of parameters is large (curse of dimensionality).

Instead of a direct calculation of $\pi(X|D;\tau)$ for entire region, the Markov chain MC (MCMC) method can be used to generate samples $X^{(t)}$ with a probability proportional to $\pi(X|D;\tau)$.
The MCMC method creates the next configuration $X^{(t+1)}$ by a small modification of the current configuration $X^{(t)}$.
The temperature $\tau$ is one of the most important hyper-parameters in MCMC sampling for the following reason.
MCMC sampling can climb over a hill with a height of $\tau$ but cannot easily escape from a valley deeper than $\tau$.
Thus, we should increase the temperature in order to avoid becoming stuck in local minima.
On the other hand, since walkers cannot see valleys smaller than $\tau$, the precision of the obtained result, $F_\text{min}$, becomes approximately $\tau$, and decreasing the temperature is necessary in order to achieve more precise results.
This dilemma indicates that we should tune the temperature carefully.

The replica-exchange MC (REMC) method~\cite{REMC-PAPER} is an extended ensemble Monte Carlo method for overcoming this problem.
In the REMC method, $K$ independent systems (replicas) with different temperatures $\tau_0, \tau_1, \dots, \tau_{K-1}$ are prepared, and the MCMC method is performed in parallel for each replica.
After some updates, $K/2$ pairs of all the replicas are created, and exchanging temperatures between replicas in each pair is attempted. This trial succeeds with some probability according to the detailed balance condition.
After the efficiently long simulation, samples with the given temperature $\tau_i$ gathered from all of the replicas follow a temperature-dependent distribution, i.e., $\pi(X|D; \tau_i)$.
In the REMC method, the temperature of each replica can change, and hence each replica can escape from local minima.

\subsection{Population-annealing Monte Carlo method \label{SEC-METHOD-PAMC}}

The population-annealing MC (PAMC) method~\cite{PAMC-PAPER} is an extension of the simulated annealing (SA) method.
The SA method starts performing the MCMC method at sufficiently high temperature and decreases the temperature gradually in order to search for the minimum of $F(X)$.
However, the temperature-decreasing process, which suddenly changes the temperature, distorts the distribution of random walkers from equilibrium.
This is because a burn-in process is needed for each temperature if the expectation values are wanted.
The annealed importance sampling (AIS) method~\cite{AIS-PAPER} compensates for this distortion by introducing an extra weight, i.e., the Neal-Jarzynskii (NJ) weight.
The AIS method prepares $N$ independent replicas and performs the SA method for these replicas in parallel.
After $N$ series of configurations are generated, the NJ weighted average of observables (e.g., $F(X)$ itself) is taken over all of the samples for each temperature.
In the AIS method, the variance of the NJ weights increases as the simulation proceeds.
This means that the effective number of samples becomes small, and, as a result, the calculation precision becomes worse.
The PAMC method introduces resampling of replicas according to their NJ weights in order to reduce the variance of weights. After each resampling step, all of the NJ weights are reset to unity.

\subsection{Comparison among the analysis methods \label{SEC-METHOD-COMPARISON}}

Finally, 
the five analysis methods are compared from the viewpoint of the application.
The Nelder-Mead method consists of iterative local updates from an initial guess and
tends to converge to a (local) minimum near the initial guess.
The Nelder-Mead method is typically useful
when an initial guess is obtained by another analysis method. 
The grid search is reduced to trivial parallel computations and is useful if the total computational cost is bearable based on the computational resources of the user. 
The grid search, however, tends to be impractical with a large data dimension $n={\rm dim}(X)$, since the number of grid points $(N_{\rm grid} \propto (N_{\rm grid}^{\rm 1D})^n)$ will be huge, as discussed, for example, in Section 4.4 of Reference \cite{HOSHI_2021_SION}. 
The BO method can realize an efficient iterative search to obtain the optimal point $X^{\rm (opt)}$ by setting a surrogate model and an initial dataset generated by a random search. 
The two MC methods generally require a large number of sampling points but give a histogram of the posterior probability density as a unique stationary state. 
The posterior probability density is crucial
when the uncertainty is estimated based on the experimental conditions and apparatus, 
as in Reference \cite{Anada_2017_JApplCrys_XRD_MC, Nagai2020} for the SXRD experiment. 

The difference between the REMC and PAMC methods appears 
in the typical number of parallel (MPI) processes $N_{\rm p}$. 
The REMC method is a parallel computation among various tolerable uncertainty parameters $\{ \tau_i \}_i$ and the typical number of possible parallel processes $N_{\rm p}$ is $N_{\rm p} =10-10^2$. 
The PAMC method is a parallel computation 
among different replicas at a given value of the tolerable uncertainty parameter $\tau = \tau_i$ and the typical number of possible parallel processes $N_{\rm p}$ is $N_{\rm p} =10^3-10^5$ or more. 
A practical comparison between the REMC and PAMC methods is ongoing on supercomputers. 

% \verb|2DMAT| enables us to use a multi-stage scheme,  
% in which
% a single analysis method is carried out first, and then other methods are preformed.
% For example, a two-stage scheme was used in Reference \cite{TANAKA_2020_Preprint},
% in which a grid search was first carried out to obtain a grid point near the minimum point, 
% and the Nelder-Mead method was then carried out to obtain a finer solution  
% with the initial guess obtained in the grid search. 
% The grid search in the first stage may be replaced by
% the BO or MC methods. 

%%%%%%%%%%%%%%%%%%%%%%%%%%%%%%%%%%%%%%
\section{Supported experiments and objective functions \label{SEC-SOLVER}}

This section is devoted to an overview of the supported experiments and problems.
\verb|2DMAT| is applicable to various experiments and problems by preparing a proper forward problem solver $D_{\rm cal}(X)$.
The present package supports 
TRHEPD, SXRD, and LEED experiments, as explained in Sections~\ref{SEC-TRHEPD}, \ref{SEC-SXRD}, and \ref{SEC-LEED}, respectively.
The present package also supports several analytic test functions for $F(X)$, and users can define a user-defined function $F(X)$, which will be explained in Sections~\ref{SEC-ANALYTIC} and~\ref{SEC-USERDEFINED}. 

\subsection{Total-reflection high-energy positron diffraction (TRHEPD)\label{SEC-TRHEPD} }

The analysis of TRHEPD data is realized, when \verb|sim-threpd-rheed| \cite{STR-URL, HANADA_PRB_1995, STR-PAPER}, an open-source simulator written in Fortran, is installed. 
Total-reflection high-energy positron diffraction is a novel experimental probe for two-dimensional materials and has been actively developed in the last decade at large-scale experimental facilities at the Slow Positron Facility (SPF), Institute of Materials Structure Science (IMSS), High- Energy Accelerator Research Organization (KEK) \cite{HUGENSCHMIDT_2016_SurfSciRep_rev, Fukaya_2018_JPHYSD, Fukaya_2019_Book_SurfaceStructure, Mochizuki_2016_PCCP_TiO2, ENDO_20208_Carbon}.
Details are provided in review studies \cite{HUGENSCHMIDT_2016_SurfSciRep_rev, Fukaya_2018_JPHYSD, Fukaya_2019_Book_SurfaceStructure}.

%%%%%%%%%%%%%%%%%%%%%%%%%%%%%%%%%%%%%%%%%%%%%%%%%%%%%%%%%%
\subsection{Surface X-ray diffraction \label{SEC-SXRD} }

The analysis of SXRD data is realized when \verb|sxrdcalc| \cite{SXRDCALC-URL,SXRDCALC-PAPER}, an open-source simulator written in the C language, is installed.
Surface X-ray diffraction has been used for decades for structure analysis experiments involving two-dimensional materials. 
Details are provided in review studies \cite{FEIDENHANSL_1989_SXRD_review,Tajiri_2020_SXRD_review}.
It is noted that several other SXRD  simulators, like 
\verb|ANA-ROD|
\cite{ANA-ROD-URL, ANA-ROD-PAPER} and
\verb|CTR-structure|
\cite{CTR-structure-URL, Anada_2017_JApplCrys_XRD_MC, Nagai2020}, are available online. 

%%%%%%%%%%%%%%%%%%%%%%%%%%%%%%%%%%%%%%%%%%%%%%%%%%%%%%%%%%
\subsection{Low-energy electron diffraction \label{SEC-LEED}}

The analysis of LEED data is realized by installing \verb|SATLEED| 
\cite{VANHOVE-1993}, 
a well-known simulator written in Fortran, 
as the forward problem solver $D_{\rm cal}(X)$.
Low-energy electron diffraction has been used for decades 
for structure analysis experiments involving two-dimensional materials.
Details are provided in review studies \cite{VANHOVE-1993}.
Low-energy electron diffraction data was analyzed by 
SATLEED in numerous studies such as Ref.~\cite{HIRAHARA-2017}. 
Note that the LEED simulator is also applicable to 
low-energy positron diffraction (LEPD) \cite{TONG2000_LEPD,Wada2018_LEPD},
recently developed 
for structure determination of two-dimensional materials.

\subsection{Analytic test functions \label{SEC-ANALYTIC}}

Several well-known analytic test functions,  
such as the Rosenbrock function \cite{Rosenbrock1960}, 
the Ackley function \cite{Ackley1987}, and 
the Himmerblau function \cite{Himmelblau_1972} 
are also implemented as the objective function $F(X)$
in order to demonstrate the algorithms 
in Section \ref{SEC-ALGORITHM}.
For example, the Himmelblau function is given as 
\begin{equation} 
F(x,y)=(x^2 + y - 11)^2 +(x+y^2-7)^2, \label{EQ-HIMMELBLAU}
\end{equation}
and the minimum value of $F=0$ appears at $(x,y) \approx (3.00, 2.00), (-2.81, 3.13), (-3.78, -3.28)$, and $(3.58,-1.85)$. 

\subsection{User-defined objective functions \label{SEC-USERDEFINED}}
Users can implement and analyze any objective function $F(X)$.
For example, in a previous study \cite{KOHASHI_2021_PerfPred}, the REMC algorithm in \verb|2DMAT| was used 
as the performance prediction method for 
a massively parallel numerical library 
for a generalized eigenvalue problem.
This paper shows a method by which to predict the elapsed time $T$ for the numerical library
as a function of the number of nodes used $P$ ($T=T(P)$) and
focuses on extrapolation to larger values of $P$.
The teacher dataset is the set of measured elapsed times for various 
numbers of nodes $\{T_{\rm exp}(P_i)\}_{i=1,...,\nu}$. 
The present paper proposed a model of the elapsed time 
$T_{\rm cal}(P,X)$ with a parameter set $X=(C_1, C_2, ..C_n)$.
The objective function to be minimized is the relative error 
\begin{equation} 
F(X)= \sum_i^\nu \frac{|T_{\rm cal}(P_i,X)-T_{\rm exp}(P_i)|^2}{|T_{\rm exp}(P_i)|^2}.
\end{equation}

%%%%%%%%%%%%%%%%%%%%%%%%%%%%%%%%%%%%%%%
\begin{figure*}[t]
\begin{center}
  \includegraphics[width=\textwidth]{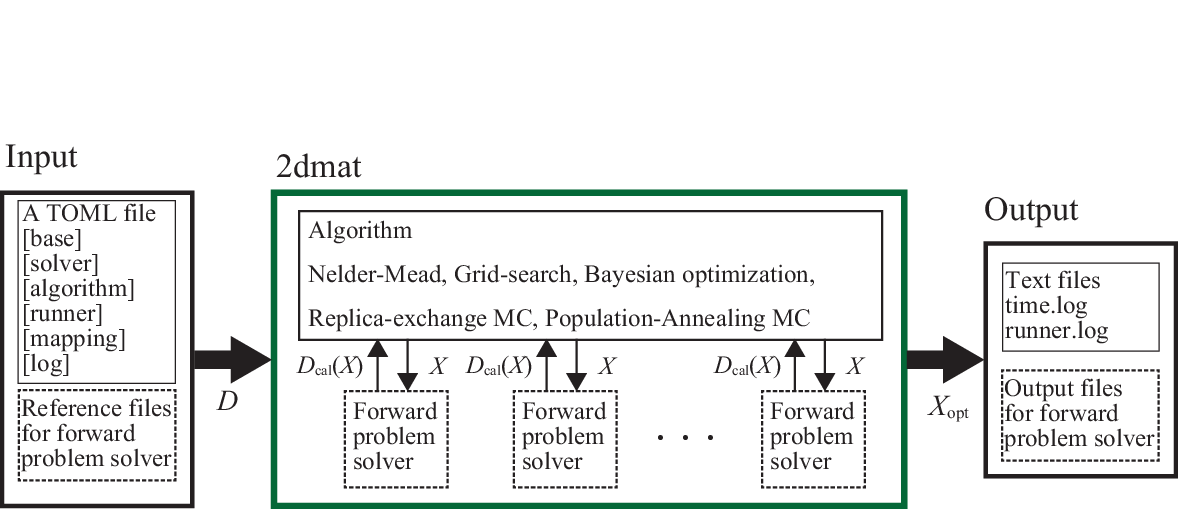}
\end{center}
\caption{
Schematic calculation flow for 2DMAT. The data to be fitted, $D$, are given by the input file. When the target quantity $X$ is given, the forward problem solver calculates the numerical data $D_{\rm cal}(X)$. 
Through an analysis algorithm, a candidate solution $X$ is proposed to minimize $|D-D_{\rm cal}(X)|$. 
When the analysis process is finished, the optimal point $X_{\rm opt}$ is output. 
In the MC method, the posterior probability density can be output as a histogram.}
\label{FIG-2dmat}
\end{figure*} %%%%%%%%%%%%%%%%%%%%%%%%%%%%%%%%%%%%%%%

%%%%%%%%%%%%%%%%%%%%%%%%%%%%%%%%%%%%%%
\section{Software \label{SEC-SOFT}}

\verb|2DMAT| is a framework for applying a search algorithm to a forward problem solver $D_{\rm cal}(X)$ in order to find the optimal solution of $F(X)$, such as determining the optimized atomic positions. 
A prototype package of \verb|2DMAT| was developed as a private program \cite{TANAKA2020_ACTA_PHYS_POLO, TANAKA_2020_Preprint,HOSHI_2021_SION}.
%for TRHEPD experiments.
We then added several new features to the code and 
released 2DMAT as an open-source software framework.
The structure of \verb|2DMAT| is shown in Fig. \ref{FIG-2dmat}. 

\verb|2DMAT| requires one input file with the TOML format~\cite{TOML} and reference files for the forward problem solver, for example, the experimental data to be fitted. The input file has the following six sections\footnote{A section, called a 'table' in TOML, is a set of parameters specified in the \texttt{name = value} format.}:

\begin{enumerate}
    \item The \verb|[base]| section specifies the basic parameters. 
    
    \item The \verb|[solver]| section specifies the parameters related to the forward problem solver.

    \item The \verb|[algorithm]| section specifies the parameters related to the algorithm.
    
    \item The \verb|[runner]| section specifies the parameters of Runner, which bridges the Algorithm and the Solver. 
    
    \item The \verb|[mapping]| section defines the mapping from a parameter searched by the Algorithm to a parameter used in the forward problem solver.
    
    \item The \verb|[log]| section specifies parameters related to logging of solver calls. 
\end{enumerate}
In addition, reference files such as the target data $D$ to be fitted and basic input files for forward problem solver are needed. After preparing these files, the optimized parameters $X_{\rm opt}$ can be searched for by running \verb|2DMAT|. 

After finishing the calculation, \verb|2DMAT| outputs the \verb|time.log|, which describes the total calculate time, and \verb|runner.log|, which contains log information about solver calls.
In addition, output files for each forward solver and the algorithm are saved.
For details, see the manual for \verb|2DMAT|.
Below, the installation and usage of \verb|2DMAT| are introduced through a demonstration.

\subsection{Install}\label{subsec:install}
\verb|2DMAT| is written in Python and requires Python 3.6.8 or higher.
Installing \verb|2DMAT| requires the python packages tomli~\cite{URL-tomli} and numpy~\cite{URL-numpy}.
In addition, as optional packages, 
scipy~\cite{URL-scipy}, and physbo~\cite{PHYSBO} are required 
the Nelder-Mead method, and Bayesian optimization, respectively.
If users want to perform a parallel calculation, mpi4py~\cite{mpi4py2021} is also required.
These packages are registered in PyPI, which is a public repository of Python software.
Since \verb|2DMAT| is also registered in PyPI, users can install \verb|2DMAT| simply by typing the following command:
\begin{verbatim}
$ python3 -m pip install py2dmat
\end{verbatim}
The sample files can be downloaded from the official site for \verb|2DMAT|.
Using git, all files will be downloaded by typing the following command:
\begin{small}
\begin{verbatim}
$ git clone https://github.com/issp-center-dev/2DMAT
\end{verbatim}
\end{small}
\red{Recently, some of authors have developed the MateriApps Installer\cite{MAInstaller}, a scripting tool that provides a computing environment for using various software packages. Users can also be used to install} \verb|2DMAT| \red{with setting the necessary libraries}.

Note that forward problem solvers used in py2dmat, such as \verb|sim-trhepd-rheed| and \verb|sxrd|, must be installed separately when users want to use these solvers.
Information on each solver, available from official websites, is in the \verb|2DMAT| manual.

\subsection{Usage}
Let us introduce the use of \verb|2DMAT| through a sample demonstration.
The input file for this sample, \verb|input.toml|, is located in the \verb|sample/analytical/bayes| directory and shown as follows:
\begin{verbatim}
[base]
dimension = 2
output_dir = "output"

[algorithm]
name = "bayes"
seed = 12345

[algorithm.param]
max_list = [6.0, 6.0]
min_list = [-6.0, -6.0]
num_list = [61, 61]

[algorithm.bayes]
random_max_num_probes = 20
bayes_max_num_probes = 40

[solver]
name = "analytical"
function_name = "himmelblau"
\end{verbatim}

In the \verb|[base]| section, the dimension of parameters to be optimized and the output directory are specified by variables \verb|dimension| and \verb|output_dir|, which are set as $2$ and \verb|output|, respectively. 
In the \verb|[algorithm]| section, the search algorithm and the seed of the random number generator are specified by parameters \verb|names| and \verb|seed|, which are set as \verb|"bayes"| and $12,345$, respectively.
There are subsections depending on the search algorithm.
In the case of selecting the algorithm \verb|"bayes"|, there are \verb|param| and \verb|bayes| subsections.
In the \verb|param| subsection, the search region can be specified by setting the maximum and minimum values, \verb|max_list| and \verb|min_list|, and the specified space is equally divided by the parameter \verb|num_list|.
In this example, the two-dimensional space [$-6, 6$], [$-6, 6$] for the $61$ divided parameters is searched.
In the \verb|bayes| subsection, the number of random samplings to be taken before Bayesian optimization \verb|random_max_num_probes| and the number of steps to perform Bayesian optimization \verb|bayes_max_num_probes| are specified.
The default score function (acquisition function) is set as Thomson sampling.
If the user wants to use expected improvement (EI) or probability of improvement (PI) as the score, then the parameter \verb|score| is available (e.g., \verb|score = "EI"|).
In the \verb|[solver]| section, the forward problem solver is specified as the \verb|himmelblau| function,
Eq.~(\ref{EQ-HIMMELBLAU}), 
as defined in the \verb|analytical| solver using \verb|name| and \verb|function_name|. 

After preparing the input file, the calculation will be performed in two ways.
The first is the \texttt{py2dmat} command, which is installed via \texttt{pip install}.
Here, \texttt{py2dmat} takes one argument, the name of the input file, as follows:
\begin{verbatim}
$ py2dmat input.toml
\end{verbatim}
The second is to use the python script \verb|py2dmat_main.py|, which is located in the \verb|src| directory.
The command for execution is given as follows (here, we assume that the current directory is \verb|sample/analytical/bayes|):
\begin{small}
\begin{verbatim}
$ python3 ../../../src/py2dmat_main.py input.toml
\end{verbatim}
\end{small}
The only difference between these methods is which python package will be used. 
The former uses the installed python package, and the latter uses the source files located in \texttt{src/py2dmat}.
After finishing the calculation, the following files are output to the \verb|output| directory. In \verb|BayesData.txt|, step $i$, the best parameter $X_{\rm best}$, and $F(X_{\rm best})$, as well as the action $X_i$, and $F(X_i)$ at the $i$-th step are output as shown below.

\begin{verbatim}
#step x1 x2 fx x1_action x2_action fx_action
0 -2.4 -0.7999999999999998 ...
1 -2.4 -0.7999999999999998 ...
...
57 3.6000000000000014 -1.7999999999999998 ...
58 3.6000000000000014 -1.7999999999999998 ...
59 3.6000000000000014 -1.7999999999999998 ...
\end{verbatim}
In this case, we can see that the optimal parameter at step $60$ is $(3.6, -1.8)$ around $(3.58, -1.85)$, one of the exact minimum points of $F(X)$ defined in Eq.~(\ref{EQ-HIMMELBLAU}).

% We will show another example of applying the PAMC method for the Himmelblau function in Section \ref{subsec:ex_him}.
% See the \verb|2DMAT| manual for more information.

%%%%%%%%%%%%%%%%%%%%%%%%%%%%%%%%%%%%%%
% \section{Examples with an analytic test function and user-defined problem \label{SEC-EXAMPLES-ANALYTIC}}

% This section shows examples with an analytic test function and a user-defined problem in order to show the general applicability of \verb|2DMAT|. 

% %%%%%%%%%%%%%%%%%%%%%%%%%%%%%%%%%%%%%%
% \subsection{Population-annealing Monte Carlo method  for the Himmerblau function} \label{subsec:ex_him}

\begin{figure}[t]
\begin{center}
  \includegraphics[width=0.45\textwidth]{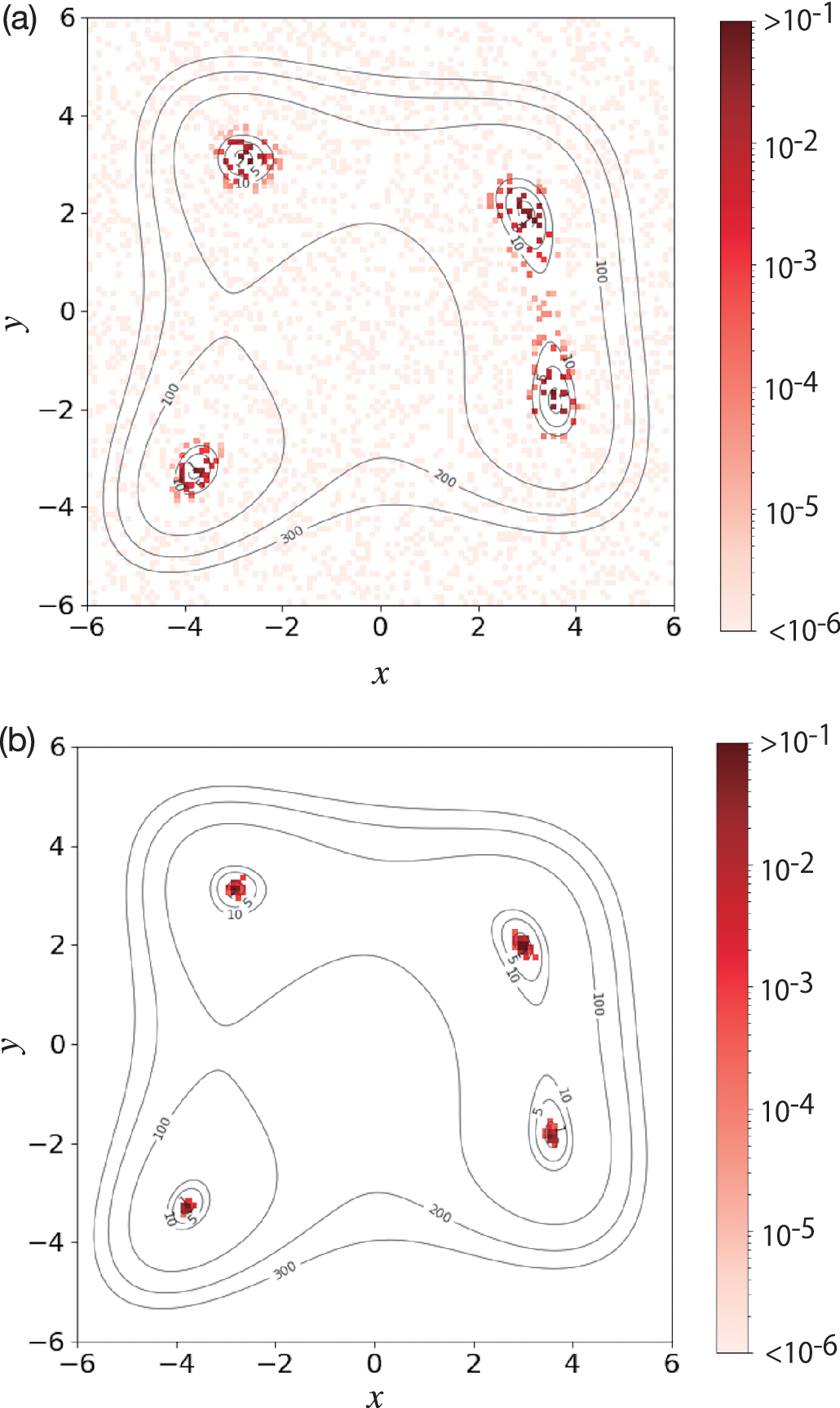}
\end{center}
\caption{Population-annealing Monte Carlo (PAMC) analysis of Himmelblau function. 
Histograms of probability density 
$\pi(x, y;\tau)$ are shown 
at (a) $\tau^{-1}  \approx 0.53$ and 
(b) $\tau^{-1}  \approx 10$. 
The contour of $F$ is also plotted as a guide for the eye. 
}
\label{FIG-HIMMELBLAU}       % Give a unique label
\end{figure}

We can easily apply another algorithm to solving the same inverse problem by editing \verb|[algorithm]| section in the input file. For example, PAMC method can be applied when we change the \verb|[algorithm]| section in the input file as follows:
\begin{verbatim}
[algorithm]
name = "pamc"
seed = 12345

[algorithm.param]
max_list = [6.0, 6.0]
min_list = [-6.0, -6.0]
unit_list = [0.05, 0.05]

[algorithm.pamc]
bmin = 0.0
bmax = 10.0
Tnum = 20
Tlogspace = false
numsteps_annealing = 20
nreplica_per_proc = 130
resampling_interval = 4
\end{verbatim}
In this case, the number of replicas was set to $N_{\rm replica}=3,120$, and 
the number of annealing steps or the number of temperature values was set to $K=20$.
The initial and final temperatures were 
$\tau_0^{-1} = 0$ and $\tau_{K-1}^{-1} = 10$, respectively.
The intermediate temperatures $\{\tau_j \}_{i=1,..,K-2}$
were prepared to discretize $[\tau_0^{-1}, \tau_{K-1}^{-1}]$ uniformly, i.e.,
\begin{equation}
\tau_j^{-1} = \tau_0^{-1} + (\tau_{K-1}^{-1} - \tau_0^{-1} ) \frac{j}{K-1}. 
\end{equation}
At each temperature $(\tau=\tau_j)$, 
$N_{\rm MCMC}^{\rm (tmp)}$=20 MCMC steps were carried out. 
The total number of MCMC steps for each replica
was $K \times N_{\rm MCMC}^{\rm (tmp)} = 20 \times 20 = 400$.
The resampling process occurred every four annealing steps. 
\red{Figure \ref{FIG-HIMMELBLAU} shows the resultant histogram of the probability density at (a) $\tau^{-1} = \tau_1^{-1} \approx 0.53$ and 
(b) $\tau^{-1} = \tau_{19}^{-1} = 10$. 
The histograms in Figures \ref{FIG-HIMMELBLAU}(a) and \ref{FIG-HIMMELBLAU}(b) 
were drawn with a bin width of $h_{\rm bin}=0.1$. 
Figure \ref{FIG-HIMMELBLAU}(b) confirms that the PAMC analysis detects the four global minima correctly. }
For details of the algorithms and their parameter settings, please refer to the \verb|2DMAT| official manual.

%%%%%%%%%%%%%%%%%%%%%%%%%%%%%%%%%%%%%%%%%%%%%%%%%%
%%%%%%%%%%%%%%%%%%%%%%%%%%%%%%%%%%%%%%%%%%%%%%%%%%

%%%%%%%%%%%%%%%%%%%%%%%%%%%%%%%%%%%%%%
\subsection{User-defined objective function}
The simplest way to define a user-defined objective function $F(X)$ is to add the function to the \verb|analytical| solver, which is for the benchmark test.
This solver is defined in the \texttt{src/py2dmat/solver/analytical.py} file, and hence a new problem can be added simply by editing this file.
When adding the function, the function can be analyzed by invoking the \texttt{src/py2dmat\_main.py} script.

Another way is to use the \verb|function| solver, which can register a python function as $F(X)$.
The following simple script (\verb|sample/user_function/simple.py|) shows an example for a grid search of a user-defined objective function $F(X) = \sum_i^N X_i^2 / N$:
\begin{lstlisting}[language=Python,numbers=left,numbersep=3pt,basicstyle=\ttfamily\footnotesize]
import numpy as np
import py2dmat
import py2dmat.util.toml
import py2dmat.algorithm.mapper_mpi as pm_alg
import py2dmat.solver.function

def my_objective_fn(x: np.ndarray) -> float:
    return np.mean(x * x)

file_name = "input.toml"
inp = py2dmat.util.toml.load(file_name)
info = py2dmat.Info(inp)
solver = py2dmat.solver.function.Solver(info)
solver.set_function(my_objective_fn)
runner = py2dmat.Runner(solver, info)
alg = pm_alg.Algorithm(info, runner)
alg.main()
\end{lstlisting}
At line 14, \texttt{solver.set\_function} registers a function \texttt{my\_objective\_fn} as the objective function.
In this way, the original source code of Py2dmat does not need to be edited.

%%%%%%%%%%%%%%%%%%%%%%%%%%%%%%%%%%%%%%%%%%%%%%%%%%
\begin{figure}[t]
\begin{center}
  \includegraphics[width=0.33\textwidth]{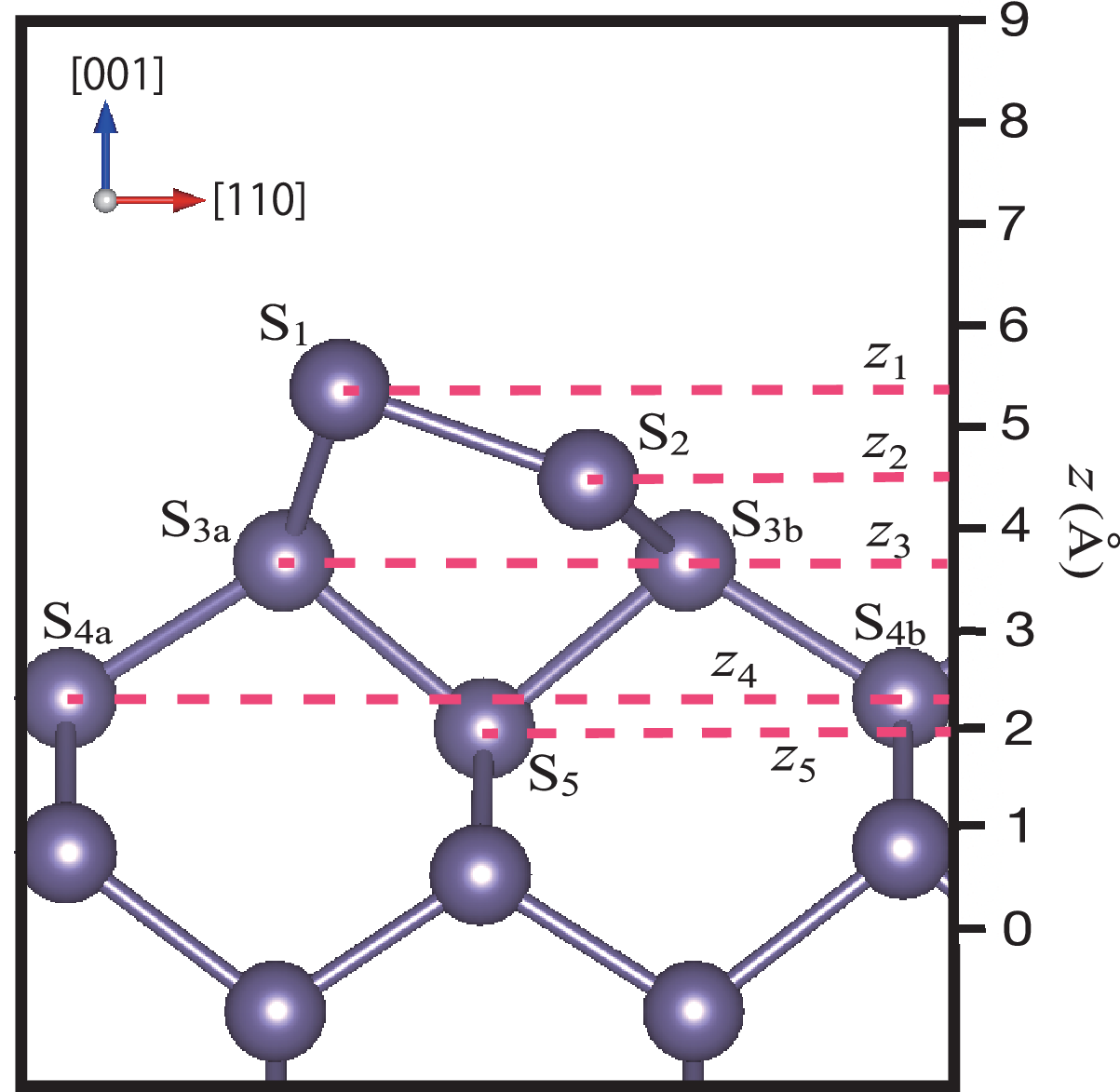}
\end{center}
\caption{
 Side view of the Ge(001)-c($4 \times 2$) surface structure. 
The $z$ coordinates of the first, second, third, \red{fourth and fifth} surface layers
are denoted as $z_1$, $z_2$, $z_3$,  $z_4$ and $z_5$, respectively. 
}
\label{FIG-Ge001-STRUCTURE}       % Give a unique label
\end{figure}
%%%%%%%%%%%%%%%%%%%%%%%%%%%%%%%%%%%%%%%%%%%%%%%%%%

%%%%%%%%%%%%%%%%%%%%%%%%%%%%%%%%%%%%%%%%%%%%%%%%%%
\begin{figure}[t]
\begin{center}
  \includegraphics[width=0.39\textwidth]{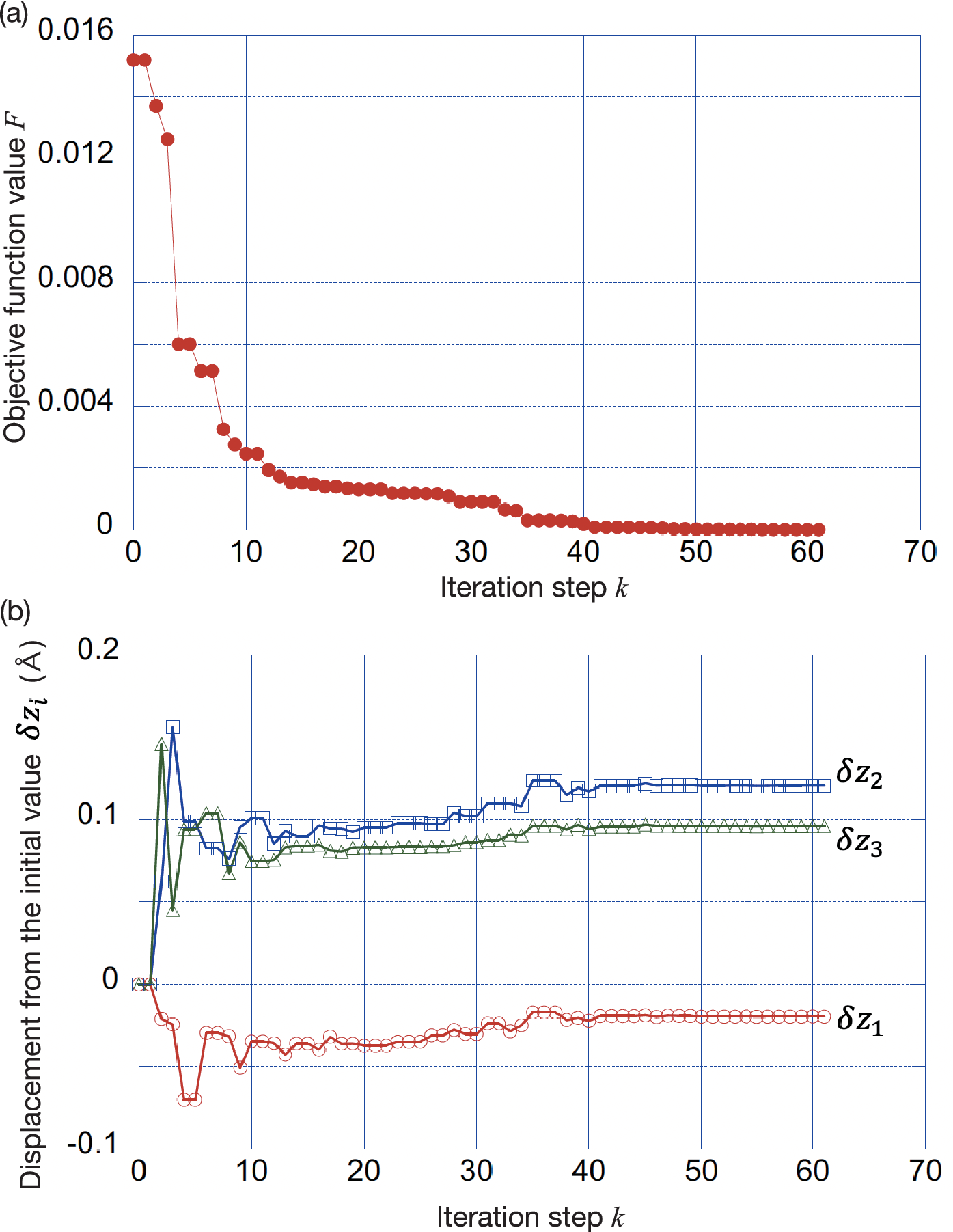}
\end{center}
\caption{
Nelder-Mead analysis of TRHEPD experiment for Ge(001)-c($4 \times 2$) surface.
(a) The objective function $F=F(z_1^{(k)},z_2^{(k)},z_3^{(k)} )$
and (b) the deviations from the initial values 
$\delta z_i^{(k)}=z_i^{(k)}-z_i^{(0)}$ ($i=1,2,3$) are plotted  as functions of 
the iteration step $k(=0,1,2...)$.}
\label{FIG-Ge001-NM-METHOD}       % Give a unique label
\end{figure}
%%%%%%%%%%%%%%%%%%%%%%%%%%%%%%%%%%%%%%%%%%%%%%%%%%

%%%%%%%%%%%%%%%%%%%%%%%%%%%%%%%%%%%%%%
\section{Examples of TRHEPD data analysis \label{SEC-EXAMPLES}}

%\subsection{Examples of TRHEPD experiment \label{SEC-EXAMPLES}}

%\subsection{Ge(001)-c($4 \times 2$) surface}
\subsection{Analysis of artificial TRHEPD data}
In this subsection, all of the analysis algorithms in Section \ref{SEC-ALGORITHM}, except for the PAMC algorithm, are demonstrated with an example of the artificial generated TRHEPD data.
The application of the PAMC method is skipped in this section, because the PAMC and REMC methods calculate, commonly, the posterior probability density $\pi(X|D; \tau_i)$, and the resultant figures for the PAMC method are almost the same as for the REMC method.
% The software manual contains an example of the TRHEPD experiment in the PAMC algorithm. 
% The software manual also contains examples of the SXRD and LEED experiments.

The present example for the TRHEPD experiment is 
the Ge(001)-c($4 \times 2$) surface, a famous semiconductor surface. 
Hereinafter, the $z$ axis is defined as being normal to the material surface.
Figure \ref{FIG-Ge001-STRUCTURE} shows a side view of the surface structure. 
The horizontal view is found, for example, in Fig. 3(b) of Reference \cite{TANAKA_2020_Preprint}.
The first and second surface atoms are denoted by S$_1$ and S$_2$, respectively, and form an asymmetric dimer.
The third surface atoms are denoted by S$_{\rm 3a}$ and S$_{\rm 3b}$.
\red{The fourth surface atoms are denoted by S$_{\rm 4a}$ and S$_{\rm 4b}$.
The fifth surface atom is denoted by S$_{\rm 5}$.
The $z$ coordinates of the first layer (S$_1$), 
the second layer (S$_2$), the third layer (S$_{\rm 3a}$ and S$_{\rm 3b}$),
the fourth layer (S$_{\rm 4a}$ and S$_{\rm 4b}$) and 
the fifth layer (S$_{\rm 5}$)
are denoted as $z_1, z_2$, $z_3$, $z_4$ and $z_5$, respectively. }

The experimental data are set to be those for the one-beam condition\cite{ICHIMIYA_1987_SurfSciLett_OneBeam, Fukaya_2018_JPHYSD}, the experimental condition for the determination of the $z$ coordinates of the atom positions.
The present analysis was carried out with artificial reference data, instead of real experimental data, as in the previous paper \cite{TANAKA_2020_Preprint}.
The reference data were generated by a simulation with reference positions of 
$(z_1, z_2, z_3) = (z_1^{\rm(ref)}, z_2^{\rm(ref)}, z_3^{\rm(ref)}) \approx (5.231 \text{\AA}, 4.371 \text{\AA}, 3.596 \text{\AA})$~
\cite{SHIRASAWA2006-LEED-Si001-Ge001,TANAKA_2020_Preprint}.

The objective function is expressed as $F = F(z_1, z_2, z_3)$.
When the values of $z_1$ and $z_2$ are replaced with each other, the objective function $F(z_1, z_2, z_3)$ is unchanged $(F(z_2, z_1, z_3) = F(z_1, z_2, z_3))$, owing to the symmetry of the structure.
Therefore, we do not lose generality by assuming the constraint $z_1 \ge z_2$.

%%%%%%%%%%%%%%%%%%%%%%%%%%%%%
\subsubsection{Nelder-Mead analysis \label{SEC-EXAMPLE-NM}}

The Nelder-Mead method was applied to 
the analysis of the Ge(001)-c($4 \times 2$) surface in a previous study \cite{TANAKA_2020_Preprint}, in which the initial values were chosen to be  
$(z_1, z_2, z_3) = (z_1^{(0)}, z_2^{(0)}, z_3^{(0)}) \equiv (5.25 \text{\AA}, 4.25 \text{\AA}, 3.50 \text{\AA})$.
Figure \ref{FIG-Ge001-NM-METHOD}(a) shows that the objective function $F$ decreases monotonically through the iterative optimization steps $k(=0,1,2,\dots)$ 
and reaches the convergence point $(z_1, z_2, z_3) = (z_1^{\rm(ref)}, z_2^{\rm(ref)}, z_3^{\rm(ref)})$ after $k=61$ iterative steps. 
Figure \ref{FIG-Ge001-NM-METHOD}(b) shows the deviations from the initial values 
$\delta z_i^{(k)}=z_i^{(k)}-z_i^{(0)}$ ($i=1,2,3$)  as functions of 
the iteration step $k$.
It can be seen that $\delta z_i^{(k)}$ is small $(|\delta z_i^{(k)} |\le 0.15 \AA)$ for the converged structure.
The files for the calculation are stored in 
the \verb|sample/py2dmat/minsearch/| directory. 

%%%%%%%%%%%%%%%%%%%%%%%%%%%%%
\subsubsection{Grid-search analysis \label{SEC-EXAMPLE-GRID}}

%%%%%%%%%%%%%%%%%%%%%%%%%%%%%%%%%%%%%%%%%%%%%%%%%%
\begin{figure}[t]
\begin{center}
  \includegraphics[width=0.37\textwidth]{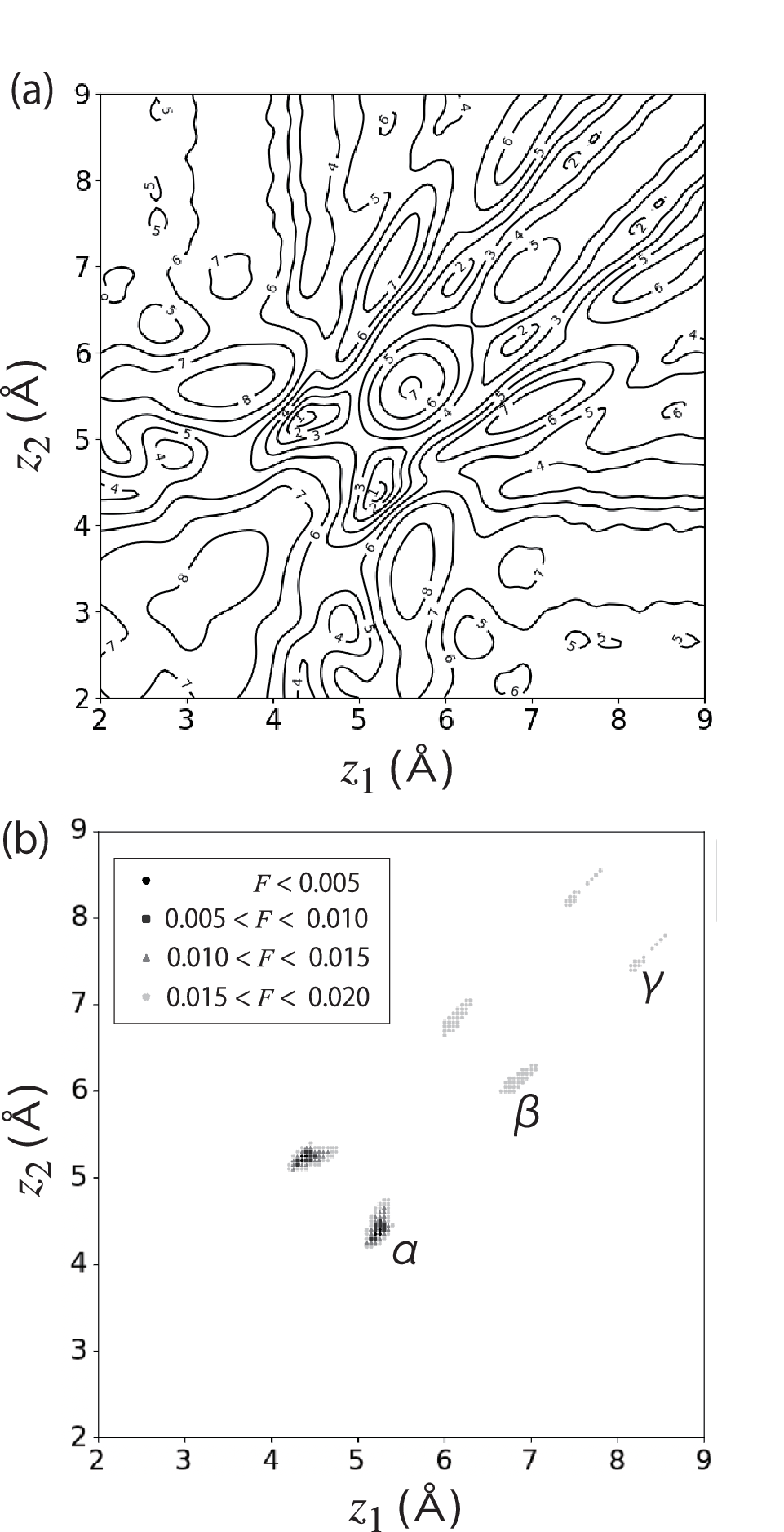}
\end{center}
\caption{
Grid-search analysis of TRHEPD experiment for Ge(001)-c($4 \times 2$) surface. 
(a) Contour plot of the objective function $F(z_1, z_2)$ 
using the grid-search data. The isovalues are shown as $100 \times F$.
(b) Objective function $F(z_1, z_2)$ on the selected grid data
that satisfy $F<0.005$ (circles), $0.005 < F < 0.010$ (squares), 
$0.010 < F < 0.015$ (triangles), and
$0.015 < F < 0.020$ (crosses). 
}
\label{FIG-GRID-CALC}       % Give a unique label
\end{figure}
%%%%%%%%%%%%%%%%%%%%%%%%%%%%%%%%%%%%%%%%%%%%%%%%%%

A grid search was applied to 
the analysis of the Ge(001)-c($4 \times 2$) surface in a previous study \cite{TANAKA_2020_Preprint},
in which the grid points were on a $(z_1, z_2, z_3)$ space 
with a grid interval of $h=0.25$ \AA. 

Figure \ref{FIG-GRID-CALC} shows the results of the grid search on a $(z_1,z_2)$ space 
with a grid interval of $h=0.05$ \AA, where $z_3$ is fixed at $z_3=z_3^{\rm(ref)}$. 
The grid range is $L_{\rm min}\equiv 2 \text{\AA} \le z_1, z_2$ $\le L_{\rm max}\equiv 9 \text{\AA}$, 
and the number of the grid points on the $z_1$ or $z_2$ axis is $N_{\rm grid}^{\rm (1D)} = (L_{\rm max}-L_{\rm min})/h+1 = 141$.
The total number of grid points is $N_{\rm grid} = (N_{\rm grid}^{\rm (1D)})^2 =141^2 = 19,881$. 
As a result, the grid minimum points are obtained
$(z_1,z_2) = (5.25 \text{\AA}, 4.4 \text{\AA}), (4.4 \text{\AA}, 5.25 \text{\AA})$ that give the minimum value of $F=0.003$.
The grid minimum point $(z_1, z_2)=(5.25 \text{\AA}, 4.4 \text{\AA})$ is close to the true minimum point
 $(z_1^{\rm(ref)}, z_2^{\rm(ref)}) \approx (5.231 \text{\AA}, 4.371 \text{\AA})$.
To obtain a finer solution than the grid minimum point, the Nelder-Mead method, in which the initial point is chosen to be the grid minimum point, can be used.

Figure \ref{FIG-GRID-CALC}(a) plots the contour of $F$ using the grid search data  
and indicates that the objective function $F(z_1, z_2)$ has many local minima. 
Figure \ref{FIG-GRID-CALC}(b) picks out 
the grid points that satisfy the condition $F<0.02$, because these points are candidate solutions in practical data analysis \cite{Fukaya_2018_JPHYSD,HOSHI_2021_SION}.
Three local regions that contain (local) minima 
are found within the condition $z_1 \ge z_2$ and 
are denoted as $\alpha$, $\beta$, and $\gamma$. 
The region $\alpha$ contains the grid minimum point $(z_1, z_2)=(5.25 \text{\AA}, 4.4 \text{\AA})$ with $F \approx 0.003$, 
while the regions $\beta$ and $\gamma$ contain local minima with $0.01 < F < 0.02$.  

Note that the three regions, $\alpha$, $\beta$, and $\gamma$, are almost linearly distributed
and satisfy the relation $( z_1 - z_2 ) \approx (\text{constant})$, which is a common property of TRHEPD.
Similar results have also been obtained for the Si$_4$O$_5$N$_3$ / 6H-SiC(0001)-($\sqrt{3} \times \sqrt{3}$) R30$^\circ$ surface, 
as shown in Fig. 3(a) of Reference \cite{HOSHI_2021_SION}. 
This stems from the fact that the interaction of positron waves scattered from the first and second atomic layers contributes significantly to the diffraction signal, and the objective function $F(z_1, z_2)$ is quite sensitive to the distance between the first and second atomic layers $(z_1 - z_2)$~\cite{HOSHI_2021_SION}. 

%%%%%%%%%%%%%%%%%%%%%%%%%%%%%%%%%%%%%%
\subsubsection{Bayesian optimization analysis}

%%%%%%%%%%%%%%%%%%%%%%%%%%%%%%%%%%%%%%%%%%%%%%%%%%
\begin{figure}[t]
\begin{center}
  \includegraphics[width=0.47\textwidth]{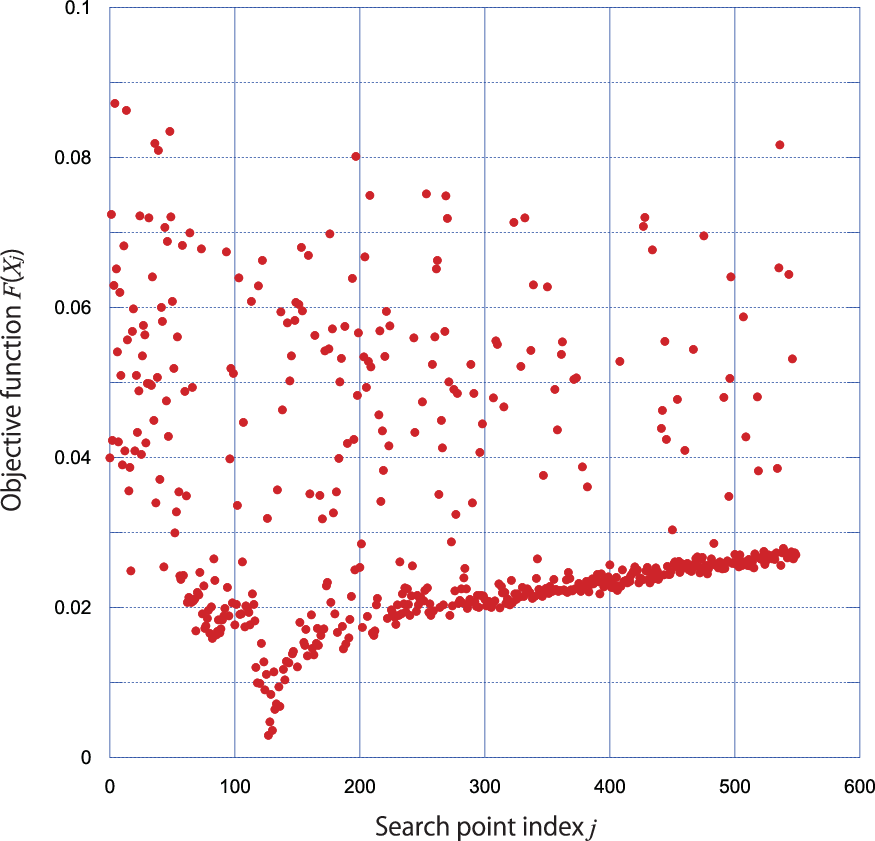}
\end{center}
\caption{
Objective function at
search points $F(z_1^{(j)},z_2^{(j)})$ with $j=0,1,2,\dots, 549$
in Bayesian optimization (BO) analysis of TRHEPD data for
Ge(001)-c($4 \times 2$) surface. 
}
\label{FIG-BO-R-Factor}       % Give a unique label
\end{figure}
%%%%%%%%%%%%%%%%%%%%%%%%%%%%%%%%%%%%%%%%%%%%%%%%%%

%%%%%%%%%%%%%%%%%%%%%%%%%%%%%%%%%%%%%%%%%%%%%%%%%%
\begin{figure}[t]
\begin{center}
  \includegraphics[width=0.40\textwidth]{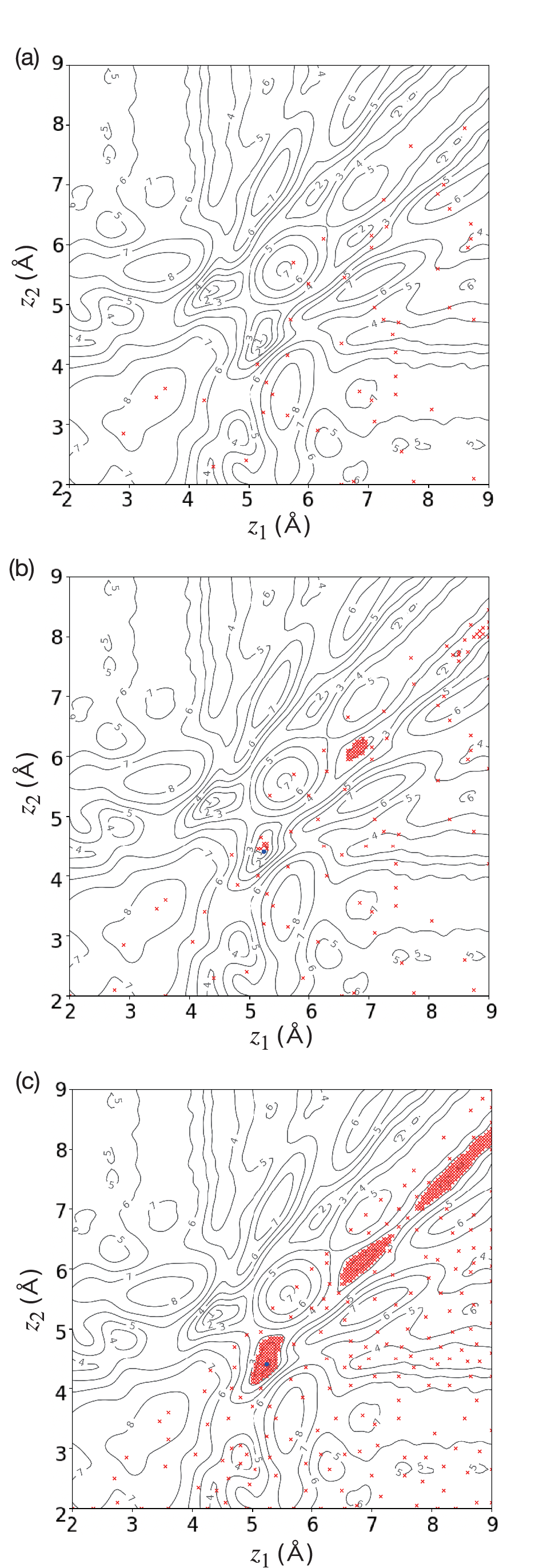}
\end{center}
\caption{
Search points $\{ (z_1^{(j)},z_2^{(j)})\}_{j}$
in BO analysis of TRHEPD data for
Ge(001)-c($4 \times 2$) surface 
with (a) $j=0,\dots,49$, (b) $j=0,\dots,127$,
and (c) $j=0,\dots,449$.
The grid minimum point $(z_1,z_2)=(5.25 \text{\AA}, 4.4 \text{\AA})$ is indicated by a filled circle, and
the other search points are indicated by crosses. 
The contour of $F$  is also plotted as a guide for the eye.
}
\label{FIG-BO-SNAP}       % Give a unique label
\end{figure}
%%%%%%%%%%%%%%%%%%%%%%%%%%%%%%%%%%%%%%%%%%%%%%%%%%

The BO analysis was carried out using a similar grid to that in Section \ref{SEC-EXAMPLE-GRID}, but
the grid points were prepared only under the constraint $z_1 \ge z_2$. 
Thus, the total number of grid points is 
$N_{\rm grid, BO} = (N_{\rm grid}^{\rm (1D)})(N_{\rm grid}^{\rm (1D)}+1)/2 =141\times 142/2 = 10,011$.

Figure \ref{FIG-BO-R-Factor} plots
the objective function $F(z_1^{(j)}, z_2^{(j)})$ as a function of the search point index $j=0,1,2,\dots,549$. 
The initial $n_{\rm rand}=50$ search points ($j=0, \dots, 49$) were generated using a random search. 
Then, the search points with $j=50, 51, \dots$ were generated iteratively using the BO algorithm. 
Figure \ref{FIG-BO-R-Factor} indicates that the grid minimum point $F$($z_1$=5.25 \AA, $z_2$=4.4 \AA)=0.003 appears at  $j$=127.
The number of search points $n_{\rm opt}=127$ is much smaller than the total number of grid points $N_{\rm grid, BO} = 10,011$ ($n_{\rm opt} \ll N_{\rm grid, BO} $), which shows the efficiency of the BO method. 

Figure \ref{FIG-BO-SNAP} shows the search points $(z_1^{(j)}, z_2^{(j)})$ that appear in the BO analysis with (a) $j \le 49$, (b) $j \le 127$, and (c) $j \le 549$. 
Figure \ref{FIG-BO-SNAP} indicates that the search points cover the region $\beta$ at first, as seen in Figure \ref{FIG-BO-SNAP}(b), and then
cover the regions $\alpha$ and $\gamma$ later, as shown in Fig. \ref{FIG-BO-SNAP}(c).

%%%%%%%%%%%%%%%%%%%%%%%%%%%%%%%%%%%%%%
\subsubsection{Replica-exchange Monte Carlo analysis}

%%%%%%%%%%%%%%%%%%%%%%%%%%%%%%%%%%%%%%%%%%%%%%%%%%
\begin{figure}[t]
\begin{center}
  \includegraphics[width=0.50\textwidth]{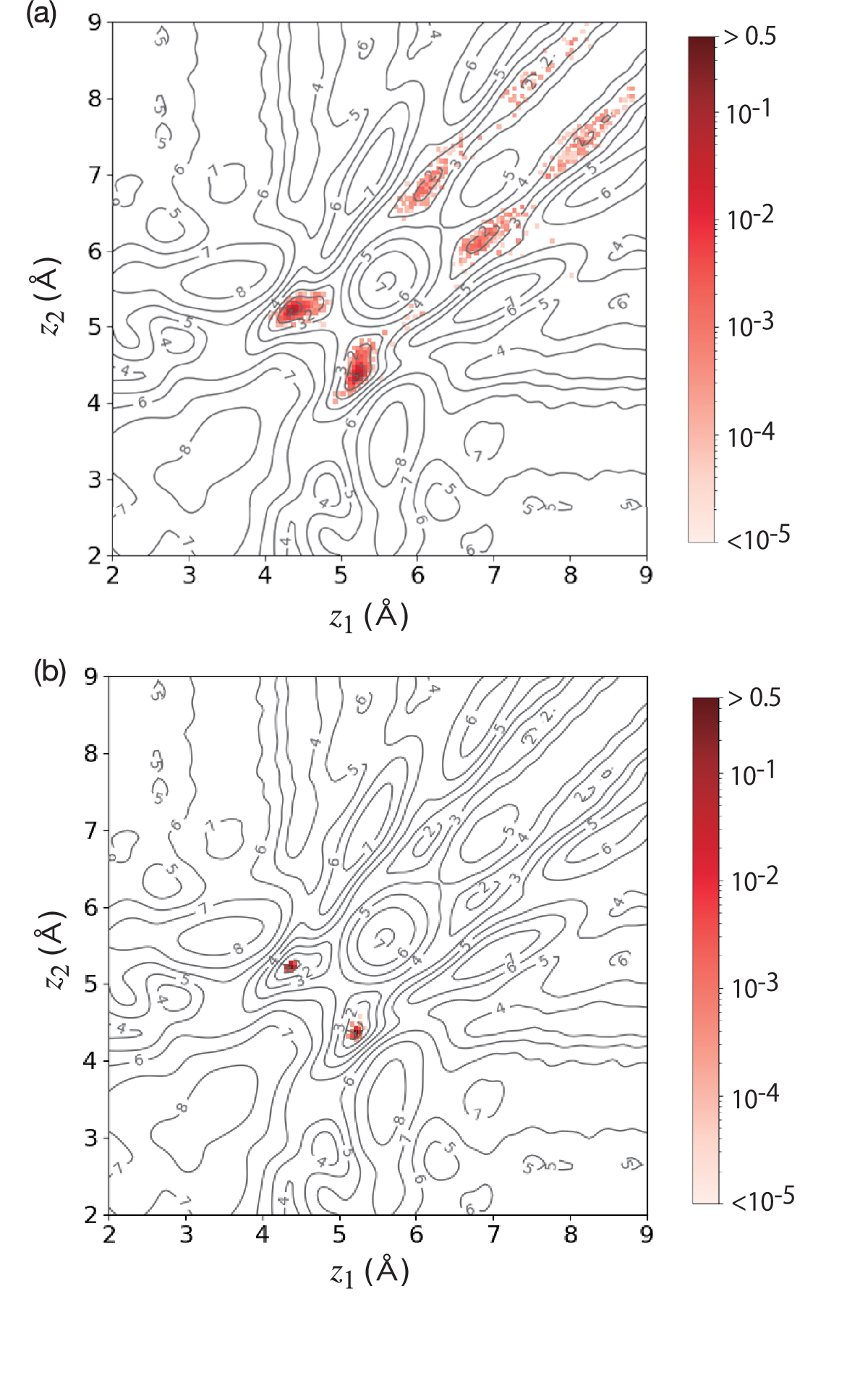}
\end{center}
\caption{Replica exchange Monte Carlo (REMC) analysis of TRHEPD data for
Ge(001)-c($4 \times 2$) surface.
The histograms of the posterior probability density 
$\pi(z_1,z_2|D; \tau)$
are drawn for temperature parameters of
(a) $\tau=0.0029$ and 
(b) $\tau=0.001$.
The contour of $F$ is also plotted as a guide for the eye.
}
\label{FIG-REMC}       % Give a unique label
\end{figure}
%%%%%%%%%%%%%%%%%%%%%%%%%%%%%%%%%%%%%%%%%%%%%%%%%%

%%%%%%%%%%%%%%%%%%%%%%%%%%%%%%%%%%%%%%%%%%%%%%%%%%
\begin{figure}[t]
\begin{center}
  \includegraphics[width=0.40\textwidth]{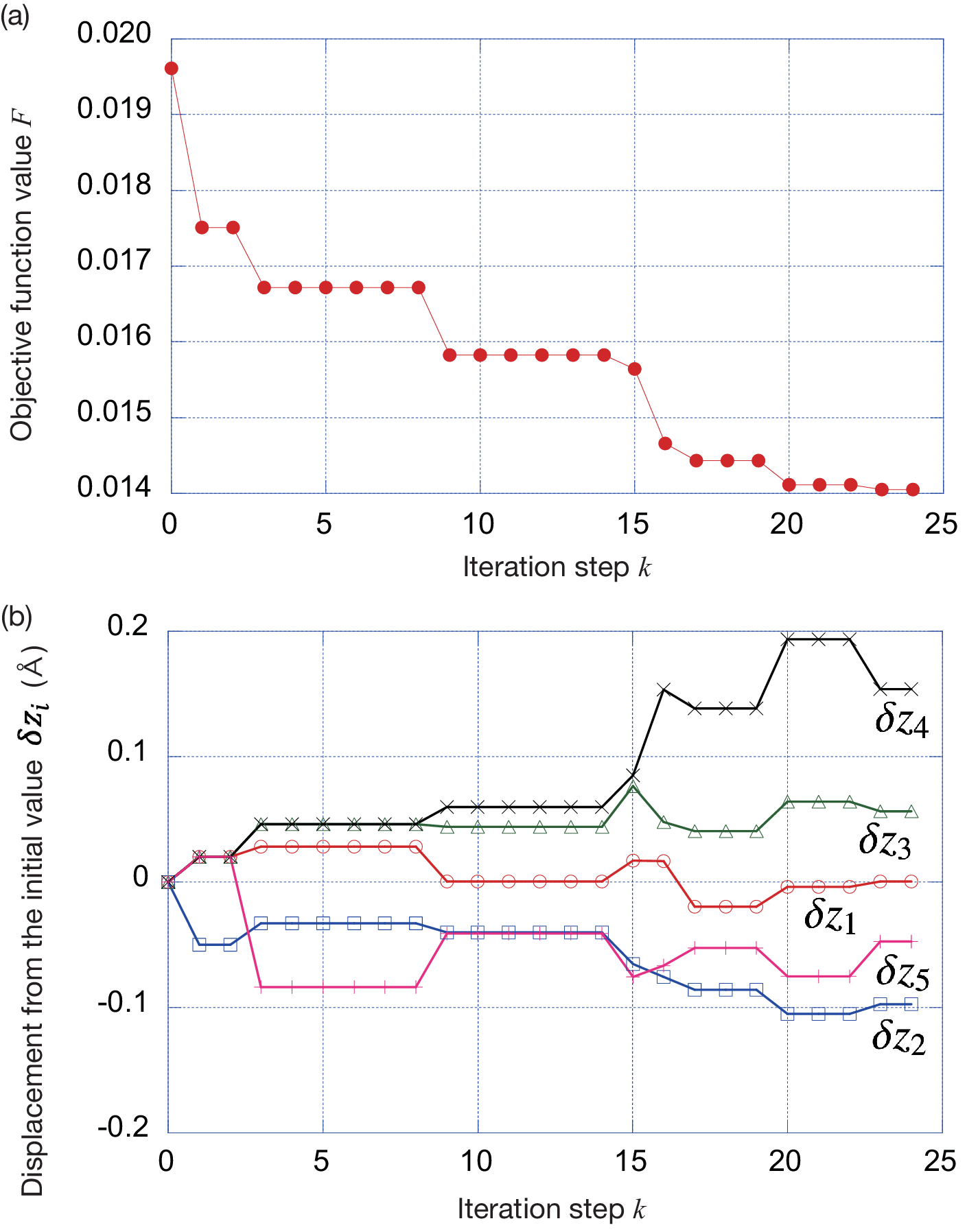}
\end{center}
\caption{
\red{Nelder-Mead optimization analysis of 
the experimental TRHEPD data for Ge(001)-c($4 \times 2$) surface.
(a) The objective function $F=F(z_1^{(k)},z_2^{(k)},z_3^{(k)},z_4^{(k)} ,z_5^{(k)}  )$
and (b) the deviations from the initial values 
$\delta z_i^{(k)}=z_i^{(k)}-z_i^{(0)}$ ($i=1,2,5$) are plotted  as functions of 
the iteration step $k(=0,1,2...)$.}
}
\label{FIG-GE001-EXP-OPT}       % Give a unique label
\end{figure}
%%%%%%%%%%%%%%%%%%%%%%%%%%%%%%%%%%%%%%%%%%%%%%%%%%

%%%%%%%%%%%%%%%%%%%%%%%%%%%%%%%%%%%%%%%%%%%%%%%%%%
\begin{figure}[t]
\begin{center}
  \includegraphics[width=0.40\textwidth]{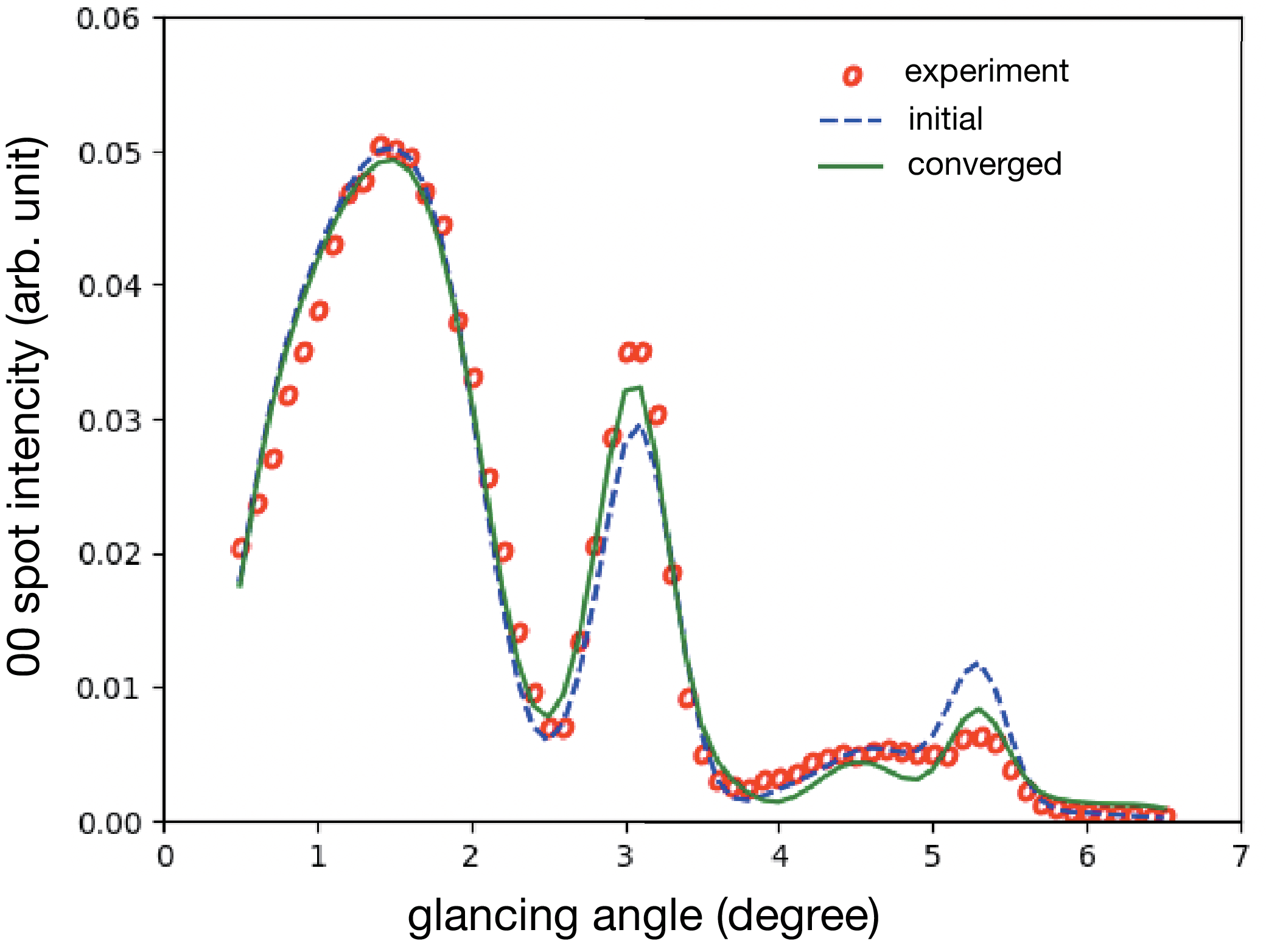}
\end{center}
\caption{\red{Comparison between the experimental rocking curve (open circles), 
the calculated ones for 
the  initial  (dashed line) and converged (solid line) structures 
in the Nelder-Mead optimization analysis of 
the experimental TRHEPD data for Ge(001)-c($4 \times 2$) surface.}
}
\label{FIG-GE001-EXP-RC}       % Give a unique label
\end{figure}
%%%%%%%%%%%%%%%%%%%%%%%%%%%%%%%%%%%%%%%%%%%%%%%%%%

Figure \ref{FIG-REMC} shows the results of the REMC method.
In this example, we used $K=36$ replicas and prepared 
datasets for temperature $\{ \tau_j \}_{j=0,..,K-1}$ 
as follows.
The maximum and minimum temperatures were 
$\tau_0 = 0.1$ and $\tau_{K-1} = 0.001$, respectively,
and the other values $\{ \tau_j \}_{j=1,..,K-2}$
were prepared to discretize $[\tau_0, \tau_{K-1}]$ uniformly on a log scale, i.e.,
\begin{equation}
\tau_j = \tau_0 \times\left(\frac{\tau_{K-1}}{\tau_0}\right)^{\frac{j}{K-1}}.
\end{equation}
The total number of MCMC steps was $N_{\rm MCMC}$ = 50,000 for each replica.
The temperature-exchange process occurred every 50 MCMC steps.
Figure \ref{FIG-REMC} shows the posterior distributions $\pi(z_1, z_2|D; \tau)$, 
as histograms, at
(a) $\tau = \tau_0 = 0.1$, (b) $\tau = \tau_{27} \approx 0.0029$ and (c) $\tau = \tau_{35} = 0.001$.
The histograms were constructed  without the \lq burn-in' data 
at the early $N_{\rm MCMC}/2$ MCMC steps and 
were drawn with a bin size of $h_{\rm bin}=$0.05 \AA. 
 The histogram in Fig. \ref{FIG-REMC}(b) 
indicates that the objective function has two minima.
In Figs. \ref{FIG-REMC}(b) and \ref{FIG-REMC}(c), 
the histogram can be confirmed to be symmetric 
$\pi(z_1,z_2|D;\tau) \approx \pi(z_2,z_1|D;\tau)$ 
owing to the symmetry of the objective function $(F(z_1,z_2)=F(z_2,z_1))$.

\subsection{\red{Analysis of experimental TRHEPD data}}

\red{
The present subsection is devoted to the analysis of the experimental TRHEPD data.
Several studies for other materials are ongoing. }

\subsubsection{\red{Ge(001)-c($4 \times 2$) surface}}

\red{
Very recently, one of the authors (I. M.) carried out the TRHEPD experiment for Ge(001)-c($4 \times 2$) surface. 
The sample preparation and the measurement condition are explained as follows. The sample substrate (5 $\times$ 10 $\times$ 0.5 mm$^3$) was cut from a mirror-polished Ge(001) wafer. The surface was cleaned by a few cycles of 0.5 keV Ar$^+$ sputtering 
($\sim$ 5 × 10$^{-4}$ Pa, 673 K, 15 min)
followed by annealing (853 K, 10 min) in a TRHEPD measurement chamber \cite{HOSHI_2021_SION, Mochizuki_2016_PCCP_TiO2, ENDO_20208_Carbon},
after which cleanliness and well-ordered periodicity of the surface were confirmed by reflection high-energy electron diffraction patterns. The TRHEPD experiment was performed using a positron beam of 10 keV. The incident azimuth was set at 22.5$^\circ$ off the [$1\bar{1}0$] direction ({\it i.e.}, the one-beam condition \cite{ICHIMIYA_1987_SurfSciLett_OneBeam,Fukaya_2018_JPHYSD}). 
The glancing angle was varied from 0.5$^\circ$ to 6.5$^\circ$.}

\red{
The experimental data was analyzed  
by the Nelder-Mead method using} \verb|2DMAT| \red{.
The iterative procedure was performed 
for the five variables $\{z_i\}_{i=1,5}$ in Fig. \ref{FIG-Ge001-STRUCTURE}, 
until the objective function $F=F(z_1, z_2, z_3, z_4, z_5)$ converges 
within the criteria $ \Delta F =5 \times 10^{-4}$.
The initial structure was set to be the one determined 
by LEED experiment \cite{SHIRASAWA2006-LEED-Si001-Ge001}.}
\red{Figure \ref{FIG-GE001-EXP-OPT}(a)  shows 
the objective function $F^{(k)}=F(z_1^{(k)}, z_2^{(k)}, z_3^{(k)}, z_4^{(k)}, z_5^{(k)})$ 
as the function of the iterative step $k(=0,1,...,24)$.
As a result, 
the objective function $F^{(k)}$ 
decreases from $F^{(0)}=1.96 \times 10^{-2}$ at the initial structure into
$F^{(24)}=1.40 \times 10^{-2}$ at the converged structure. 
Figure \ref{FIG-GE001-EXP-OPT}(b)  shows 
the deviation of the variables $\{z_i\}_{i=1,5}$ from the initial values 
$\delta z_i^{(k)} = z_i^{(k)} - z_i^{(0)}$
as the function of the iteration step $k$.
Here one finds that the difference between the initial and final structures 
was approximately $0.15$ \AA~ or less. 
In other words, the structures determined by the TRHEPD and LEED experiments agree with a small (0.1\AA-scale) difference. 
Figure \ref{FIG-GE001-EXP-RC}  shows 
the calculated rocking curves 
in the initial and converged structures, 
together with the experimental data. 
The calculated rocking curves in the converged structure 
agrees with the experimental data better than that of the initial structure.}

\red{
It is noted that,
among existing TRHEPD researches  
(Refs.~\cite{Fukaya_2018_JPHYSD, HOSHI_2021_SION}
and references therein),  
a structure is accepted as a final solution, 
when the objective function value is  less than 0.02 ($F \le 0.02$).
The present result demonstrates that 
the TRHEPD diffraction data is sensitive to 
a small (0.1\AA-scale) difference in the coordinates.  } 

\subsubsection{\red{Si$_4$O$_5$N$_3$ / 6H-SiC(0001)-($\sqrt{3} \times \sqrt{3}$) R30$^\circ$ surface}}

\red{
A previous paper ~\cite{HOSHI_2021_SION} reports that} 
\verb|2DMAT| \red{was used in 
the data analysis of TRHEPD experiment
for Si$_4$O$_5$N$_3$ / 6H-SiC(0001)-($\sqrt{3} \times \sqrt{3}$) R30$^\circ$ surface, a novel two-dimensional semiconductor.
The trial atomic positions were optimized by 
the Nelder-Mead method, in which
the value of the objective function
decreases 
from $F=2.28 \times 10^{-2}$ into $F=0.91 \times 10^{-2}$. 
In addition, 
a grid search was carried out 
for a local region near the optimized position,
so as to investigate the properties of 
the objective function in the local region.}

% \section{Future Utilities}

% The 2DMAT package is  
% a general-purpose data-analysis framework and
% several application studies beyond those in this paper 
% are ongoing and several types of the studies are picked out below; 
% The first type of the ongoing studies 
% is the the analysis of 
% the experimental data of SXRD and LEED. 
% In particular,
% 2DMAT will enable us a fruitful analysis,
% when we perform multiple types of the diffraction experiments on a specimen and analyze the results seamlessly.
% The second type of the ongoing studies 
% is the prediction of material properties,
% which is realized by {\it ab initio} calculations 
% with the determined atomic positions. 
% For this purpose, 
% some of the present authors developed a tool
% that converts 
% the atomic position data file used in TRHEPD analysis
% into the file of Quantum Espresso, 
% a famous {\it ab initio} software package \cite{STR-PAPER}.
% The third type of the ongoing studies 
% is an application in other problem beyond 2D materials,
% where we construct a user-defined problem,
% as explained in Sec. \ref{SEC-USERDEFINED}.

%%%%%%%%%%%%%%%%%%%%%%%%%%%%%%%%%%%%%%
\section{Summary and \red{future utilities}\label{SEC-SUMMARY}}

A general-purpose open-source software framework \verb|2DMAT| has been developed for analysis of experimental data using an inverse problem approach. 
%Therefore, \verb|2DMAT| can be used to perform large-scale and highly accurate calculations.
The framework provides five analysis algorithms: the Nelder-Mead optimization, the grid search, the Bayesian optimization, the replica-exchange Monte Carlo, and the population-annealing Monte Carlo methods. Thus, users can choose an appropriate analysis method according to the purpose and can construct a multi-stage scheme, in which a single method is carried out first, and then other methods are performed. 
The current version of \verb|2DMAT| supports analysis of several experiments; TRHEPD, SXRD, and LEED data by offering adapters for the external software packages for simulating these experiments. 
\verb|2DMAT| is designed to define the objective function by users.
Thus, by adding the appropriate objective function, many experimental measurements can be analyzed using \verb|2DMAT|. 

\red{Finally, we comment on the future utilities of} \verb|2DMAT|\red{. Several application studies beyond those in this paper  are ongoing and several types of the studies are picked out below: The first type of the ongoing studies is the the analysis of the experimental data of SXRD and LEED. In particular, }\verb|2DMAT|\red{will enable us a fruitful analysis, when we perform multiple types of the diffraction experiments on a specimen and analyze the results seamlessly. The second type of the ongoing studies is the prediction of material properties, which is realized by {\it ab initio} calculations  with the determined atomic positions. For this purpose, some of the present authors developed a tool that converts the atomic position data file used in TRHEPD analysis into the file of Quantum Espresso, a famous {\it ab initio} software package} \cite{URL-qe,STR-URL, STR-PAPER}\red{. The third type of the ongoing studies is application to other problems beyond 2D materials, where we construct a user-defined problem, as explained in Sec. }\ref{SEC-USERDEFINED}. \red{With these features and possible future utilities, we expect that }\verb|2DMAT| \red{becomes a useful data-analysis framework in material science and other scientific fields.}

\section*{Acknowledgement}

We would like to acknowledge support for code development 
from the ``Project for advancement of software usability in materials science (PASUMS)''
by the Institute for Solid State Physics, University of Tokyo. 
The present research was supported in part by  
a Grant-in-Aid for Scientific Research (KAKENHI) from the Japan Society for the Promotion of Science 
(19H04125, 20H00581). 
\verb|2DMAT| 
was developed using the following supercomputers: 
the Ohtaka supercomputer at the Supercomputer Center, Institute for Solid State Physics, University of Tokyo; 
the Fugaku supercomputer through the HPCI projects (hp210083, hp210267);  
the Oakforest-PACS supercomputer as part of the 
Interdisciplinary Computational Science Program in the Center for Computational Sciences, University of Tsukuba;
the Wisteria-Odyssey supercomputer at the Information Technology Center, University of Tokyo; 
and the Laurel2 supercomputer at the Academic Center for Computing and Media Studies, Kyoto University.
The present research was also supported in part by JHPCN project (jh210044-NAH). 
We would like to acknowledge Takashi Hanada, Wolfgang Voegeli, 
Tetsuroh Shirasawa, Rezwan Ahmed, and Ken Wada for fruitful discussions on the diffraction experiments, and Daishiro Sakata, Koji Hukushima, Taisuke Ozaki, and Osamu Sugino for fruitful discussions on the code.

\bibliographystyle{apsrev}
%\bibliography{mybibfile}






% \end{thebibliography}
\end{document}